\documentclass[preprint,aps]{revtex4}
\usepackage{graphicx}
\usepackage{dcolumn}
\usepackage{amsmath}
\begin{document}
\baselineskip 24 true pt   
\title{Quantum Treatment of the Anderson-Hasegawa Model -- Effects of
Superexchange and Polarons}

\author{Manidipa Mitra, P. A. Sreeram and Sushanta Dattagupta}
\affiliation{S. N. Bose National Centre for Basic Sciences, JD Block, 
Sector III, Salt Lake City, Kolkata-700098, India}
\email{mmitra@bose.res.in, sreeram@bose.res.in, sdgupta@bose.res.in}
\begin{abstract}
We revisit the Anderson-Hasegawa double-exchange model and critically 
examine its exact solution when the core spins are treated quantum mechanically.
We show that the quantum effects, in the presence of an additional 
superexchange interaction between the core spins, yield a term, the 
significance of which has been hitherto ignored. The quantum considerations
further lead to new results when polaronic effects, believed to be ubiquitous
in manganites due to electron-phonon coupling, are included. The consequence
of these results for the magnetic phase diagrams and the thermal heat capacity
is also carefully analysed.

\vspace{0.5in}

\noindent
PACS No. 63.20. K, 71.38, 75.30. E \\

\end{abstract}
\maketitle
\section{Introduction}
The emergence of manganites as a technologically important material due
primarily to the occurrence of colossal magnetoresistance (CMR) in 
e.g.  $La_{1-x}Ca_xMnO_3$ \cite{jin} has rekindled the interest of the 
condensed matter 
physics community in the double-exchange mechanism.
The physics of double-exchange has successfully correlated ferromagnetism and 
metallicity in the doping range of $0.2 < x < 0.4$. The basic ingredient
of double-exchange, proposed by Zener fifty years ago \cite{zen}, is 
encapsulated within a simple two site model of Anderson and Hasegawa 
\cite{and,degen,kubo}. The latter has led to a plethora of theoretical and 
experimental investigations in 
recent years \cite{rev1,rev2}, which have gone on to add one or the other 
feature to the original model, often without abundant care, as we shall argue 
here.

The undoped manganite system is characterised by the presence of an incomplete 
d-shell of the Manganese ions, which consists of 3 electrons in the $t_{2g}$ 
state and one electron in the $e_g$ state. The $t_{2g}$ spins are deep inside 
the d-level and are assumed to be unimportant in the process of charge 
transfer. However, these spins do show a tendency to align 
antiferromagnetically both in the parent compound and in the completely doped 
system. 
The Coulomb energy cost for the $e_g$ electrons 
to hop onto the adjacent Manganese ion in the absence of a hole is very large. 
The undoped system is thus an insulator. 

We will henceforth refer to the $e_g$ electron as the itinerant electron, 
the $t_{2g}$ electrons as the core electrons and the total spin in the 
$t_{2g}$ level as the core spin. Doping the system with
a divalent atom like $Ca$ or $Sr$ leads to creation of holes in the $e_g$ level
in a fraction of the Manganese ions. The itinerant electron from the 
occupied $e_g$ level of one Manganese site can then hop into its nearest 
neighbor Oxygen site, which facilitates hopping to the nearest neighbor 
unoccupied $e_g$ level of another Manganese ion, thus leading to a finite
conductivity. This process, called the ``double exchange mechanism", 
plus the presence of strong Hund's rule coupling between the core spin and 
the itinerant electron, results in an indirect coupling between neighboring 
core spins . This in turn relates the magnetic order of the underlying lattice 
with the kinetic energy of the itinerant electron. 

The conclusions derived by Anderson and Hasegawa \cite{and} can be summarised 
as follows. In the limit when Hund's coupling is infinitely strong, the 
itinerant electron would like to have its spin aligned with the local core
spin. Additionally, if the core spin is treated \underline {classically}, the
appropriate axis of quantization is the direction of the core spin vector, as 
far as the itinerant electron is concerned. Now, as the latter hops, it has
to readjust its spin to be realigned with the new core spin partner, amounting
to a rotation of the quantization axis by an angle $\theta$, which is the
polar angle between the core spins $\vec{S_1}$ and $\vec{S_2}$. From the 
property of the spin-1/2 rotation operator it follows
that the hopping or the overlap matrix element $t$  will be renormalized to 
$t \cos(\theta/2)$, assuming azimuthal symmetry. It then follows that
if $\theta$ equals $\pi$, the core spins have antiferromagnetic (AFM) 
coupling and hopping of the itinerant electron is totally inhibited. On the 
other hand, if $\theta$ equals zero, the core spins have ferromagnetic
(FM) coupling and hopping is accentuated, thus synergising transport with 
ferromagnetism. This is the first result of Anderson and Hasegawa. The latter
then proceeded to a quantum treatment of the core spins, but still operating
within the infinite Hund's coupling limit. Interestingly, it turns out that
the energy eigenvalues are identical to the earlier classical case, provided
$\cos(\theta/2)$ is identified as a \underline{parameter} which equals 
$(S_0+1/2)
/(2S+1)$, where $S$ = $\mid \vec{S_1}\mid$ = $\mid \vec{S_2}\mid$ and 
$S_0$ = $\mid \vec{S_1} + \vec{S_2} + \vec{\sigma} \mid$, $\vec{\sigma}$
being the spin of the itinerant electron ($\mid \vec{\sigma} \mid$ = 1/2).
Curiously, since $\vec{S_1}$, $\vec{S_2}$ and $\vec{\sigma}$ add up to 
$(2S+1/2)$ in the FM case for large Hund's coupling, the parameter
$\cos(\theta/2)$ would indeed be equal to unity, as in the classical
case. But in the AFM case, the parameter reduces to $1/(2S+1)$, which goes
to the classical value of zero only when $S\rightarrow \infty$, thus 
necessitating an additional constraint on the core spin, if the ``classical"
interpretation is to be taken seriously. Needless to say, in between FM and
AFM cases, the parameter $\cos(\theta/2)$ would go through a set of
discrete (and not continuous) values, thus pointing to the need of a more
careful treatment of the core spins in the quantum case. 

One of the directions in which the Anderson-Hasegawa treatment has been 
extended is to recognize the importance of an additional superexchange term 
between the core spins proportional to 
$\vec{S_1}.\vec{S_2}$. It has been assumed in the literature till now
that the superexchange term can be simply taken as an extra term to be added to
 the Hamiltonian and that the large Hund's rule coupling affects only the 
process
of charge transfer in these systems. In this paper we show inter alia that 
superexchange is itself modified in a nontrivial manner if the core spins
are dealt with quantum mechanically, which leads to a correlated diagonal 
disorder in these systems, even in the cleanest samples. Our starting point 
therefore is the Hamiltonian for a two site one electron model, including the 
superexchange interaction, given by

\begin{equation}
H=-t\sum_\tau (c_{1\tau }^{\dagger }c_{2\tau }+h.c.)-J_{H}\sum _{i=1}^{2}
\vec{S_{i}}.\vec{\sigma _{i}} + J \vec{S_1}.\vec{S_2}.
\end{equation}
Here $t$ is the hopping matrix element for the itinerant electron
between the two sites, $c_{i\tau }^{\dagger }$
($c_{i\tau }$) is the creation(annihilation) operator of the itinerant electron
at site $i$ having spin projection $\tau$, $J_{H}$ is the Hund's rule coupling 
strength, $\vec{S_{i}}$ is the core spin at site $i$ and 
$\vec{\sigma _{i}}$ is the spin of the itinerant electron at the site $i$.
The parameter $J$ is the superexchange interaction strength between the 
core spins in the 
nearest neighbor sites.  For our case we consider $\mid \vec S_i \mid = S$, 
i.e.  the core spins on all the sites are taken to have the same value.

With the preceding background, the motivation behind our work and the plan of 
this paper are as follows. We reiterate that the two site, single electron, double-exchange model
is the simplest basic framework for interpreting a large number of fascinating 
properties of manganites. While Anderson and Hasegawa did provide a quantum 
solution to the model, especially in the limit of large Hund's rule coupling, 
subsequent authors seem to have gone ahead in a somewhat cavalier fashion, in 
our opinion, about the classical limit of infinitely large core spins. This has
led to some confusion about the interpretation of parameters in the model which
needs to be cleared. As the value of the core spins in most studied CMR systems
is indeed finite --- three-halves for manganese --- it is important to delineate
the quantum versus classical effects, especially while considering additional 
phenomena, e.g., polaron-induced hopping and thermodynamics. With this aim in 
mind, we organize the paper as follows. We present in Sec. II, the exact 
quantum mechanical solution to the Anderson-Hasegawa model with an additional
superexchange term, take the large Hund's rule coupling limit and discuss the
outcome of a hitherto ignored `site-disorder' term. In Sec. III we reexamine
the issue of polaron-assisted hopping in the light of our fully quantum 
calculation. The results in this section are then employed
in Sec. IV for the computation of heat capacity and phase diagram, wherein we 
also specify the difference between our results and those which 
treat the core spins classically. 
Finally, in Sec. V, we present some concluding remarks.


\section{Exact solution for the Anderson-Hasegawa Model}

In order to find the ground state of the system we follow the quantum
mechanical calculation carried out by Anderson and Hasegawa \cite{and}. We 
first note that the Hund's rule coupling term, proportional to $J_H$ is 
diagonal in the states given by $\mid \psi_1^{\pm}\rangle$ = 
$\mid S_1,\frac{1}{2},(S_1 \pm \frac{1}{2}),S_2;S_0,M>$ and 
$\mid \psi_2^{\pm}\rangle$ = $\mid S_1,\frac{1}{2},S_2,(S_2 \pm 
\frac{1}{2});S_0,M>$, 
while the hopping part of the Hamiltonian connects these two sets of states, 
corresponding as it does to a recoupling of the itinerant electron's spin 
(1/2) from the site spin $S_1$ to $S_2$ and is thus given by the Wigner 6j 
(or Racah) coefficient (W)\cite{racah}. Here, 
$M$ = $S_0^z$. The superexchange term proportional to $J$, is off-diagonal in 
the basis states chosen above. However, it is diagonal in the states given 
by $\mid \phi(S^\prime) \rangle$ = $\mid \frac{1}{2},S_1,S_2,(S^\prime);S_0,M 
\rangle$, where $S^\prime$ = $\mid \vec{S_1}+\vec{S_2}\mid$. We can then relate
the states $\mid \psi_{1,2}^{\pm} \rangle$ to the states $\mid \phi(S^\prime)
\rangle$ through appropriate Racah coefficients again and we find, 
\begin{eqnarray}
\mid \psi_{1}^{\pm} \rangle &=& \sum_{S^\prime} \sqrt{\left[2(S_1+\frac{1}{2})
+1\right]} \sqrt{(2S^\prime+1)} ~W\left(\frac{1}{2} S_1 S^\prime S_2;(S_1+
\frac{1}{2})S^\prime\right) \mid \phi(S^\prime)\rangle , \nonumber\\
\mid \psi_{2}^{\pm} \rangle &=& \sum_{S^\prime} \sqrt{\left[2(S_2+\frac{1}{2})
+1\right]} \sqrt{(2S^\prime+1)} ~W\left(\frac{1}{2} S_1 S^\prime S_2;(S_2+
\frac{1}{2})S^\prime\right) \mid \phi(S^\prime)\rangle.
\label{trans-1} 
\end{eqnarray} 
Clearly, since $S^\prime$ (the total core spin) must couple to the itinerant 
electron spin to give the total angular momentum $S_0$, the only values of
$S^\prime$ to be summed over are $S^\prime$ = $S_0+\frac{1}{2}$ and $S_0-
\frac{1}{2}$ . The particular Racah coefficients which occur (with
$S_1=S_2=S$), have convenient closed expressions such as, 
\begin{eqnarray}
\mid \psi_1^+ \rangle &=& \cos(\alpha/2) \mid \phi(S_0-\frac{1}{2})\rangle + 
\sin(\alpha/2) \mid \phi(S_0+\frac{1}{2}) \rangle , \nonumber \\
\mid \psi_1^- \rangle &=& -\sin(\alpha/2) \mid \phi(S_0-\frac{1}{2})\rangle + 
\cos(\alpha/2) \mid \phi(S_0+\frac{1}{2})\rangle ,
\nonumber \\
\label{transform-2}
\end{eqnarray}
where,
\begin{equation}
\cos(\alpha/2)= \left[\frac{\left(2S+S_0+\frac{3}{2}\right)}
{2(2S+1)}\right]^{1/2}.
\end{equation}
The relations between $\mid \psi_2^{\pm}\rangle$ and $\mid \phi(S_0\pm\frac{1}
{2})\rangle$ states are the same as those between $\mid \psi_1^{\pm}\rangle$ 
and $\mid \phi(S_0\pm\frac{1}{2})\rangle$.
Note that the expression for $\cos(\alpha/2)$ can be written in terms of 
$\cos(\theta/2)$, which actually gives the relation between $\alpha$ and 
$\theta$ as $\alpha = \theta/2$.
Thus the Hamiltonian matrix in the space of $\mid \psi_1^\pm \rangle$ and 
$\mid \psi_2^\pm \rangle$ can be written as follows :
\begin{equation}
\left (
\begin{array}{cccc}
P_1 & P_2 &  -t \cos(\theta/2) & -t \sin(\theta/2) \\
&&&\\
P_2 & P_3 & t \sin(\theta/2) & -t \cos(\theta/2) \\
&&&\\
-t \cos(\theta/2) & t \sin(\theta/2) & P_1 & P_2 \\
&&&\\
-t \sin(\theta/2) & -t \cos(\theta/2) & P_2 & P_3
\end{array}
\right),
\end{equation}
where,
\begin{eqnarray}
P_1 &=& \frac{J}{2}\left(R_1 \sin^2(\alpha/2) + R_2 \cos^2(\alpha/2) \right)
        -\frac{J_H}{2} S, \nonumber \\
P_2 &=& \frac{J}{2}(R_1-R_2) \cos(\alpha/2)\sin(\alpha/2),\nonumber \\
P_3 &=& \frac{J}{2}\left(R_1 \cos^2(\alpha/2) + R_2 \sin^2(\alpha/2)\right)
        +\frac{J_H}{2} (S+1), 
\end{eqnarray}
and,
\begin{eqnarray}
R_1 = \left ( S_0+\frac{1}{2}\right ) \left (S_0 + \frac{3}{2} \right )-
2 S (S+1), \nonumber \\
R_2 = \left ( S_0-\frac{1}{2}\right ) \left (S_0 + \frac{1}{2} \right )-
2 S (S+1). 
\end{eqnarray}
The eigenvalues (E) of the Hamiltonian matrix are obtained from,
\begin{equation}
2 E= \frac{J_H}{2}+K_1(J) \pm
 \sqrt{4 t^2 + K_T^2(J) + K_3^2(J)
\pm 4 t \cos(\theta/2)\sqrt{ K_T^2(J)+ K_3^2(J)}},
\label{exact-1}
\end{equation}
where,
\begin{eqnarray}
K_1(J)&=& J\left[\left( S_0+\frac{1}{2}\right)^2-2S(S+1)\right] \nonumber \\
K_T(J)&=&\left[\frac{J_H(2S+1)}{2}+ K_2(J)\right] \\
K_2(J)&=& J\cos(\alpha) (S_0+\frac{1}{2})  \\
K_3(J)&=& J\sin(\alpha) (S_0+\frac{1}{2}) .
\end{eqnarray}
While the exact result of Eq.(\ref{exact-1}) may be of interest in its own
right, we examine the limit of large Hund's rule coupling, by expanding the 
square root term (upto $O(1/J_H)$). We find that the lowest energy eigenvalues 
are given by
\begin{equation}
E_m = -\frac{J_H S}{2}-t\cos(\theta/2)+\frac{J}{2}\left[\bar{S}^\prime
(\bar{S}^\prime+1)-2S(S+1)\right]+J\sin^2(\alpha/2)(S_0+\frac{1}{2}),
\label{emin-1}
\end{equation}
where $\bar{S}^\prime =  S_0 - 1/2$. 

The first two terms in Eq. (\ref{emin-1}) are the terms obtained by Anderson
and Hasegawa. We emphasise once again that the parameters
$\cos(\theta/2)$ and $\cos(\alpha/2)$ take \underline{ discrete} values which 
depend on the quantum values of the core spins. The third term is the result 
of the superexchange interaction in the absence of the itinerant electron. The 
fourth term, a novel one, is purely due to the double-exchange 
mechanism in the presence of the itinerant electron. The explicit form of this 
term is given by,
\begin{equation}
\Delta E_J = \frac{J}{2}\frac{2S-\bar{S}^\prime}{2S+1}(\bar{S}^\prime+1).
\end{equation}
It is to be noted that in the ferromagnetic limit (i.e.  $\bar S^\prime = 2S$) 
$\Delta E_J$ vanishes exactly.  

There are primarily three important points to be made about $\Delta E_J$ :

(a) This term is an on-site term : It does not involve physical transfer
of the itinerant electron from one site to the other.

(b) This term vanishes in the absence of the itinerant electron. Thus, it
exists only on the site at which the electron resides and hence, at any site
$i$, it will be proportional to $n_i$, where $n_i$ is the number of itinerant
electrons at the site $i$ and is taken to be either 1 or 0 in the absence of
double occupancy. 

(c) At any site $i$ this term depends on 3 spin values : the spin of the 
itinerant electron on the site $i$ ($\sigma_i$), the core spin at site 
$i$ ($S_i$) and the core spin on the neighboring site ($S_j$),
$j$ being the neighbor of the site $i$.

Thus, the extra energy term $\Delta E_J$ corresponds to a site energy term
which is correlated with the core spins of the nearest neighbors. We 
may therefore propose an effective double-exchange Hamiltonian in the full
lattice as
\begin{equation}
H_{{\rm eff}}= \sum_i \epsilon_i n_i -t \sum_{<ij>} \cos(\theta_{ij}/2) 
(c_i^\dagger c_j + H.C.) +  \sum_{<ij>} J_{ij} \vec{S_i}.\vec{S_j} ,
\end{equation}  

where
\begin{equation}
\epsilon_i = \sum_j \frac{J_{ij}}{2}\frac{2S-S^\prime_{ij}}{2S+1}
(S^\prime_{ij}+1),
\end{equation}
and
\begin{equation}
S^\prime_{ij} = \mid \vec{S_i} + \vec{S_j} \mid .
\end{equation}
We have explicitly taken $J_{ij}$ in order to accomodate effects of 
anisotropic superexchange also.

The classical limit of the extra term is given by
\begin{equation}
\Delta E_J^{Cl}  = \frac{J}{2} (2S+1)\left[1-\cos(\theta/2)\right]\cos(\theta/2)
 ,
\end{equation}
which goes to zero in both the ferromagnetic as well as the antiferromagnetic 
limits. Again, we see a clear distinction between the classical and the 
quantum results in the 
antiferromagnetic limit. We emphasise that in taking the purely classical
expression, one actually loses the effect of the quantum fluctuations which
are present in these systems not only because of the fluctuating spins but also
due to the on site disorder and the hopping, correlated with the spins
on the lattice. A variety of interesting physical phenomena could be studied
by taking into consideration the quantum Hamiltonian. One of these concerns 
the polaron effects, which are discussed below in Section III.

\section{Anderson-Hasegawa-Holstein Model}

Experiments in manganites - both thermodynamic and transport - seem to suggest
the importance of polaron formation and the consequent localization of charge 
carriers \cite{millis95}. The minimal model which reflects such lattice
carrier interaction on the double-exchange can be introduced by dovetailing
the Holstein mechanism on the Anderson-Hasegawa Hamiltonian. Therefore, in 
view of the results presented in Sec. II, in the limit of large Hund's rule
coupling, we may write a two site Anderson-Hasegawa-Holstein Hamiltonian as,

\begin{eqnarray}
H &=& \epsilon \sum_{i=1}^2 \sum_{\sigma}  n_{i \sigma} 
- \sum_{\sigma} t\frac{S_0 + \frac{1}{2}}{2 S + 1}(c_{1\sigma}^{\dag} c_{2 \sigma}+ c_{2 \sigma}^{\dag} c_{1 \sigma})
+ g_1 \omega_0  \sum_{i=1}^2\sum_{\sigma}  n_{i \sigma} (b_i + b_i^{\dag})\nonumber \\ 
&+& g_2 \omega_0  \sum_{\sigma} \left[ n_{1 \sigma} (b_{2}
+ b_{2}^{\dag}) +n_{2 \sigma} (b_1 + b_1^\dagger)\right]
+  \omega_0 \sum_{i=1}^2  b_i^{\dag} b_i 
+ J \vec S_{1}.\vec S_{2} +\Delta E_J,
\label{qdh-1}
\end{eqnarray} 
where, $g_{1}(g_{2})$ denotes the on-site (intersite) electron-phonon 
coupling strength and $\epsilon $ is the bare site energy. 
This site energy
is to be contrasted with the site energy which was derived in the previous
section. Here, since we are interested in doing a two-site model, the extra
term in the Hamilonian due to the superexchange interaction, has been
introduced simply as $\Delta E_J$.
Note that we have considered a single phonon mode for interatomic 
vibrations of frequency $\omega_0$ for which $b_i$ and $b_{i}^{\dag}$ are the 
annihilation and creation operators. 

We separate out the in-phase mode and the out-of-phase mode by introducing new 
phonon operators $a=~(b_1+b_2)/ \sqrt 2$ and $d=~(b_1-b_2)/\sqrt 2 $ in
the Hamiltonian. 
The in-phase mode does not couple to the
electronic degrees of freedom whereas the out-of-phase mode does, leading
to a Hamiltonian $H_d$, given by,
\begin{equation}
H_d = \omega_0 d^\dagger d +\epsilon \sum_{i=1}^2 n_i - t\left(\frac{S_0+\frac{1}{2}}{2S+1}\right) (c_1^\dagger c_2 + h.c.) + g_-\omega_0(n_1-n_2)(d+d^\dagger) 
+J \vec S_1.\vec S_2 + \Delta E_J,
\end{equation}
which represents an effective electron-phonon system. 
Following \cite{jayee} we use a Modified Lang-Firsov (MLF) transformation with 
variable phonon basis and obtain,

\begin{eqnarray}
\tilde{H_d}&=& e^R H_d e^{-R}\nonumber\\
&=&\omega_0  d^{\dag} d + \sum_{i} \epsilon_p n_{i} - 
t\frac{S_0 + \frac{1}{2}}{2 S + 1} ~ [c_{1}^{\dag} c_{2}~ \exp
(2 \lambda (d^{\dag}-d)) 
+ c_{2}^{\dag} c_{1}~\exp(-2 \lambda (d^{\dag}-d))] \nonumber \\
&+& \omega_0 (g_- - \lambda)(n_1 - n_2)(d + d^{\dag}) 
+ J \sum_{<ij>} \vec S_{i}.\vec S_{j} + \Delta E_J ,
\label{qdh-2}
\end{eqnarray}
where $R =\lambda (n_1-n_2) ( d^{\dag}-d)$, $\lambda$ is a
variational parameter related to the displacement of the $d$ 
oscillator, $g_{-}=(g_1-g_2)/\sqrt 2$
and $\epsilon_p = \epsilon - \omega_0 ( 2 g_{-} - \lambda) \lambda$. 
The basis set is given by 
$|\pm,N \rangle = \frac{1}{\sqrt 2} (c_{1}^{\dag} \pm  c_{2}^{\dag})$
$|0\rangle_e  |N\rangle$,    
where $|+\rangle$ and $|-\rangle$ are the bonding and the antibonding 
electronic states and $|N\rangle$ denotes the $N$th excited oscillator  
state within the MLF phonon basis. 
The diagonal part of the Hamiltonian $\tilde{H_d}$ in the chosen basis 
is treated as the unperturbed Hamiltonian ($H_0$) 
and the remaining part of the Hamiltonian $H_{1}= \tilde{H_{d}}-H_0$,  
as the perturbation. 
The unperturbed energy of the state $| \pm,N\rangle$ is given by 
\begin{eqnarray}
 E_{\pm,N}^{(0)}&=& \langle N,\pm|H_0|\pm, N \rangle\nonumber\\
&=& N \omega_0  + \epsilon_p  \mp t_{eff} \left[ \sum_{i=0}^{N}
 \frac{(2\lambda)^{2i}}{i!} (-1)^i N_{C_i}\right] +J\vec S_1 . \vec S_2
+\Delta E_J
\end{eqnarray}
where $t_{eff}=t~\frac{S_0 + \frac{1}{2}}{2 S + 1} ~\exp{(-2\lambda^2)}$,
$N_{C_i}=\frac{N!}{(N-i)!~~ i!}$.
The general off-diagonal matrix elements of 
$H_1$ between the two states $|\pm,N \rangle$ and $|\pm,M \rangle$ 
may be calculated for $(N-M)>0$ as in Ref. \cite{jayee}.

The unperturbed ground state is the $|+\rangle|0\rangle$ state 
and the unperturbed energy, $ E_0^{(0)}=\epsilon_p - t_{eff} + 
J \vec S_1.\vec S_2 + \Delta E_J $. However, in this exact quantum limit of core spins, 
for given values of $g_-$ and $J$, 
$E_0^{(0)}$ can have four values corresponding to ferromagnetic (FM), 
canted 1 (CA1), canted 2 (CA2) and antiferromagnetic (AFM)
orientation of the two spins for 
$\mid \vec S_{12} \mid = \mid \vec S_1+ \vec S_2 \mid = 3,2,1,0$ 
respectively. Minimizing the unperturbed ground state energy $\lambda $ is 
calculated and is given by

\begin{eqnarray}
\lambda&=&\frac {\omega_0g_{-}}{\omega_0+2t_{eff} }  .
\end{eqnarray}

We have evaluated the perturbation correction to the energy upto the sixth
order and the wave function upto the fifth order. The 
convergence of the perturbation series is very good for $t/\omega_0 \le 1$.
To obtain the ground state spin order of the core spins we calculate the
energy for each set of values of $g_{-}$ and $J$ with four possible 
$\vec S_{12}$ and find out the combination for which the energy is the minimum. Further, to study the effect of an 
external magnetic field ($\vec h$) we include a term 
$- \tilde g \mu_B (\vec S_1 + \vec S_2). \vec h$ to the Hamiltonian in equation
(\ref{qdh-1}), $\tilde g$ being the Lande g factor. We assume
that the external magnetic field is along the direction of $\vec S_{12}$ and
is expressed in units of $\mu_{eff}(=\tilde g \mu_B)$=1.

It is expected that the charge transfer from site `1' to `2' depends on the
spin order of the core spins as well as the electron-phonon interaction. 
In the double-exchange model, the effective
hopping reaches its maximum value in the ferromagnetic state and decreases 
as it approaches the antiferromagnetic limit. Moreover, in a lattice, 
the electron produces
lattice deformations and which in turn localize the electron for strong
electron-phonon coupling. To study the polaronic character one calculates  
the static correlation functions $\langle n_1 u_{1}\rangle_{0}$ and
$\langle n_1 u_{2}\rangle_{0}$,
where $u_1$ and $u_2$ are the lattice deformations at sites 1 and 2
respectively, produced by an electron at site 1 \cite{jayee,euro}. In the 
present report
with a two-site one electron model, following \cite{euro},
we calculate $- <n_1(u_1-u_2)>_0/g_{-}=\frac{\lambda^{corr}}{g_{-}}$ and study 
the nature of the polaron
crossover for different ranges of $g_{-}$ and $J$. In the `large' polaron
limit this parameter takes a small value, while with increasing electron-phonon
coupling it tends to unity, showing a distinct crossover from `large' to 
`small' polaron behavior. The measure of delocalization of the electron for 
various ranges of $g_{-}$ as well as $J$ will be evident from the kinetic 
energy. So we have also calculated the 
kinetic energy, given by,
\begin{equation}
t_{eff}^{KE}=-E_{Kin}=<\psi_G|
t \frac{S_0 + \frac{1}{2}}{2 S + 1}
~ [c_{1}^{\dag} c_{2}~ \exp
(2 \lambda (d^{\dag}-d))
+ c_{2}^{\dag} c_{1}~\exp(-2 \lambda (d^{\dag}-d)) ~]
|\psi_G>,
\end{equation} 
where $\psi_{G}$ is the ground state wave-function, is
evaluated upto the fifth order in the perturbation. The numerical evaluation
of $E_{Kin}$ will be presented below in Sec. IV.

\section{Phase diagrams and Specific Heat}
	Recently, there have been many experimental reports on manganites at 
low doping and low temperatures with and without an external
magnetic field \cite{cv,cv1,cv2}. Okuda et al have estimated the electronic 
specific heat for
$La_{1-x}Sr_xMnO_3$ in the ferromagnetic regime and concluded that the carrier
mass-renormalization near the metal-insulator transition at $x=0.16$ is minimal.
They have also observed a decrease in the low temperature specific heat in the
presence of a magnetic field. Motivated by these observations, we have carried
out a calculation of the specific heat, based on
the partition function of the system which, from a cumulant
expansion upto the 2nd order, is given by \cite{sd},

\begin{equation}
Z(\beta) = Z_0(\beta) exp{(-\int^{\beta}_{0} d\beta^\prime
\int^{\beta^{\prime}}_{0} d\beta^{\prime \prime}\langle \tilde H_1(\beta^\prime)\tilde H_1(\beta^{\prime \prime})\rangle)} ,
\end{equation}
where $Z_0(\beta) = Tr (e^{-\beta H_0})$ ; 
$\tilde H_{1} (\beta) = e^{\beta H_0} H_1 e^{-\beta H_0}$, and 
$\beta = \frac{1}{K_B T}$. The expression $\langle \rangle$ denotes the 
usual canonical averaging. The specific heat is then calculated (in arbitrary
units) from the well known relation:

\begin{equation}
C_V=-\frac{d}{dT}(\frac{d}{d\beta}lnZ(\beta)),
\end{equation}
and in the low temperature regime, to which only the zero-and one-phonon states
contribute.

	If the localized core spin at each site is $\frac{3}{2}$ then
the possible values of $\mid \vec S_1 + \vec S_2 \mid = S_{12}$ are 3, 2, 1 and
0. The ferromagnetic (FM) and antiferromagnetic (AFM) orders are obviously 
related to $S_{12}=$ 3 and 0, whereas $S_{12}=$ 2 and 1 are referred as
canted 1 (CA1) and canted 2 (CA2) states respectively. The Fig. 1 shows the 
phase diagram for the four possible spin orders for our
system, in the $g_{-}$ vs $J$ plane. For small values of $g_{-}$ and $J$, the 
FM 
state is the most stable one, and with increasing $J$, the ground state first
becomes CA1 and then CA2. For a very large value of the superexchange 
interaction $J$, the system is in an AFM order for any value of $g_{-}$. 
However, with 
increasing electron-phonon interaction $g_{-}$, the CA1 and CA2 phase become 
narrower. Indeed, for larger values of $g_{-}$
the FM state appears for very low $J$ but with  
a small increase of $J$ the system transits to the AFM phase. The CA1 and CA2 
phases in fact do not appear at all as the phase 
changes from the FM to AFM state with increasing superexchange interaction $J$,
for large values of $g_{-}$. It can be further shown that for a very large 
value of $g_{-}$ the ground state is AFM for any value of $J$. 

It is evident from the phase diagram that for a particular value of 
$J$ the ground state changes as the electron-phonon coupling ($g_-$) 
increases (Fig. 1). But the 
change of phase from one to another is not continuous with $g_{-}$, for the
quantum consideration of the core spins. This is shown in Fig. 2. 
For small values of $J=$ 0.01, the FM state exists even for a large value of 
$g_{-}$ 
and then it sharply changes to the AFM state. On the other hand, for larger 
values of $J=$ 0.04, 0.09, 
the system passes sharply to the canted phases (CA1 and CA2) and then the 
AFM state,
with increasing $g_{-}$. For $J=$0.04 the CA1 and CA2 regimes are very narrow
and for $J=$0.09 the CA1 and CA2 orders persist for a wider range of $g_-$.
This is to be contrasted with the classical core spin model in which a
similar study shows that the transitions to different core spin
orientations are continuous for the same range of  values for $g_{-}$ and $J$ 
\cite{euro}.
In the classical case only three phases(FM, AFM and Canted) are present. 
The relative angle $\theta$ between classical core spins can take any 
value from $0$ to $\pi$, so any spin orientation other than FM ($\theta=0$) 
and AFM ($\theta = \pi$) yields a canted phase. Hence, in the classical limit 
of the core spins, for certain values of $J$, the 
FM-AFM transition is a smooth and continuous transition with $g_{-}$, whereas 
for spin $\frac{3}{2}$, the  FM-AFM transition with $g_{-}$ is never continuous
 for any $J$.

The probability of hopping of the itinerant electron from site to site is a
maximum in the FM state as would be expected from the double-exchange 
mechanism.
But, for a very strong electron-phonon coupling $g_{-}$, the electron
may be localized forming a small polaron. For low values of $J$ we find both
small and large polaron ground states in the FM phase (Fig. 3). 
The large to small polaron crossover
is indicated by the relative deformation of the two lattice sites which is 
measured
by the static correlation function $\frac {\lambda_{corr}}{g_{-}}$. 
In Fig. 3 the kinetic energy $t_{eff}^{KE}$ is large for small values of 
$g_{-}$, where the polaron is large, and for large $g_{-}$, the kinetic energy 
reduces rapidly, while  
$\frac {\lambda_{corr}}{g_{-}}$ rises, showing a smooth crossover to the small 
polaron 
regime. The classical and quantum formulations of the double-exchange model
turn out to be the same in the FM limit of the core spins. So the nature of 
the kinetic energy and
the polaron crossover in the FM state, as shown in Fig. 3, will be unaltered in
the classical limit of core spins. However, in the AFM limit the two approaches
(classical and quantum) are not equivalent, as has been argued earlier also. In 
the $S \rightarrow \infty$ limit the
hopping probability is zero for the AFM case, while for $S=\frac{3}{2}$ the 
parameter modifying the hopping probability $t$, takes a finite value $0.25$ 
resulting in a finite charge transfer, even in the AFM limit. 
 
In Fig. 4 we show the nature of variation of the kinetic energy as well as the
polaron crossover in different magnetic ground states. For $J=0.09$ the ground
state is FM for low $g_{-}$, and with increasing $g_{-}$, the ground state 
changes sharply to CA1, CA2 and lastly to the AFM state. 
Since at each transition (from FM 
$\rightarrow $ CA1 $\rightarrow $ CA2 $\rightarrow $ AFM) the effective hopping 
reduces due to the double-exchange interaction, it is obvious that the kinetic 
energy will show a sharp drop at each transition point. It is expected that
the polaron crossover will also show concomitant sharp jumps at each magnetic 
transition
and the crossover to small polaron behavior will occur at lower value of 
$g_{-}$ than in the
FM limit. This is shown clearly in Fig. 4. It is further evident that in a 
double-exchange system, both the magnetic transitions and the electron-phonon 
coupling localize the
electron, when the polaron crossover and magnetic transitions are overlapping.
The locations of the large polaron region (A) and the small polaron region (B) 
are indicated in the $g_-$ vs $J$ phase diagram (Fig. 1). In Fig. 1, for large 
values of $g_-$, the
CA1 and CA2 phases are very narrow and appear as a single phase boundary
of FM-AFM region.
So the line of separation of polaronic regimes (A and B) appears as a point 
for CA1 and CA2 in Fig. 1. Thus the results presented in Fig. 3 and Fig. 4 have
a bearing on the transport behavior of our model. 

Having discussed transport we now redirect our attention to thermodynamic
properties. With this in mind we show in Fig. 5 the variation of the specific 
heat in the low temperature region in the FM state with zero and one phonon 
states. With application of an external magnetic field $\vec h$, 
$C_V$ takes lower values than for $\vec h= 0$ which is expected, as the average 
energy decreases with application of $\vec h$ in the FM state. 
For CA1($\mid \vec S_{12}\mid=2$), CA2($\mid \vec S_{12}\mid=1$) and AFM 
($\mid \vec S_{12} \mid = 0$) states the external magnetic field will tend 
to align the
core spins to feromagnetic order($\mid \vec S_{12}\mid = 3$). For CA1, CA2 and 
AFM states at low field and low temperatures
it can be shown from the present calculation that $C_V$ does not change much
from the $\vec h=0$ limit as long as $\vec h$ does not shift 
$\mid \vec S_{12}\mid$ to higher values. 
For larger $\vec h$, as the ground state changes from lower 
$\mid \vec S_{12}\mid $ to a higher 
one, $C_V$ decreases in the low temperature region. For CMR materials, there are
some reports on measurements of $C_V$ but these are 
measured in the FM state \cite{cv, cv1, cv2}. In such cases, it was found that 
for low doping regions, $C_V$ decreases
with an increasing magnetic field. Our present calculation of $C_V$ seemingly 
agrees with these experimental findings. The difference in the quantum and
classical cases for specific heat, as far as the core spins are concerned,
is exemplified in Fig. 6 and Fig. 7 for FM and AFM cases respectively.
The quantum results evidently yields the correct low-temperature limit.

\section{Conclusions}
The Anderson-Hasegawa model, though almost fifty years old, is able to capture
all the crucial features of the double-exchange mechanism, originally 
proposed by Zener. The model is restricted to just two sites but the limitation
should not be too serious when the electron hopping, influenced by thermal 
fluctuations, lattice distortions, phonon effects and other interactions, is 
expected to be incoherent. Incoherent hopping, albeit quantum in nature, is
quite distinct from coherent band-like propagation, and approximately follows a 
Markovian process as far as the quantum diffusion of the electron is concerned.
For a Markovian process only the pre-hopping and post-hopping sites matter.
Therefore, the two site abstraction of the underlying three dimensional lattice
provides the simplest paradigm which can be exploited for analyzing a variety
of phenomena which are of current interest in manganites. With this in mind, 
we have felt the need of carefully reexamining the exact quantum solution of
the Anderson-Hasegawa model, for realistic values of the core spins. We have
further used this model as the basic building block, in order to incorporate
one or the other phenomena of relevance to manganites. These include 
superexchange and polarons, which have been the focus of our attention here. 
Indeed, we have found that Superexchange, when properly treated in conjunction
with quantum spin dynamics of the core spins, leads to additional terms in the
Anderson-Hasegawa Hamiltonian, which are absent in the classical approximation
of the core spins. 
A similar effect can also be observed with the exact
solution of the Anderson-Hasegawa model, with the addition of phonon
coupling. This may lead to the enhancement of the site energy term, which
will have important consequences in these systems, especially in the
paramagnetic state of the manganite systems.
Moreover, the discreteness associated with the effective
hopping in this quantum case was shown to have further consequence for
thermodynamic and transport properties. 
In conclusion, therefore, we find that the Anderson-Hasegawa model continues to
remain relevant for the understanding of topically important issues in 
manganites.

\section{Acknowledgement}

The authors are deeply grateful to Prof. B. Dutta Roy for very helpful 
discussions on the Racah algebra for the Anderson-Hasegawa model. Discussions
with Prof. S. Satpathy on the polaronic mechanism have been quite valuable. SD
wishes to thank Prof. S. D. Mahanti for generating initial interest in 
manganites.

\newpage

\vskip 1.8in

{\bf Figure Captions : }

\noindent 

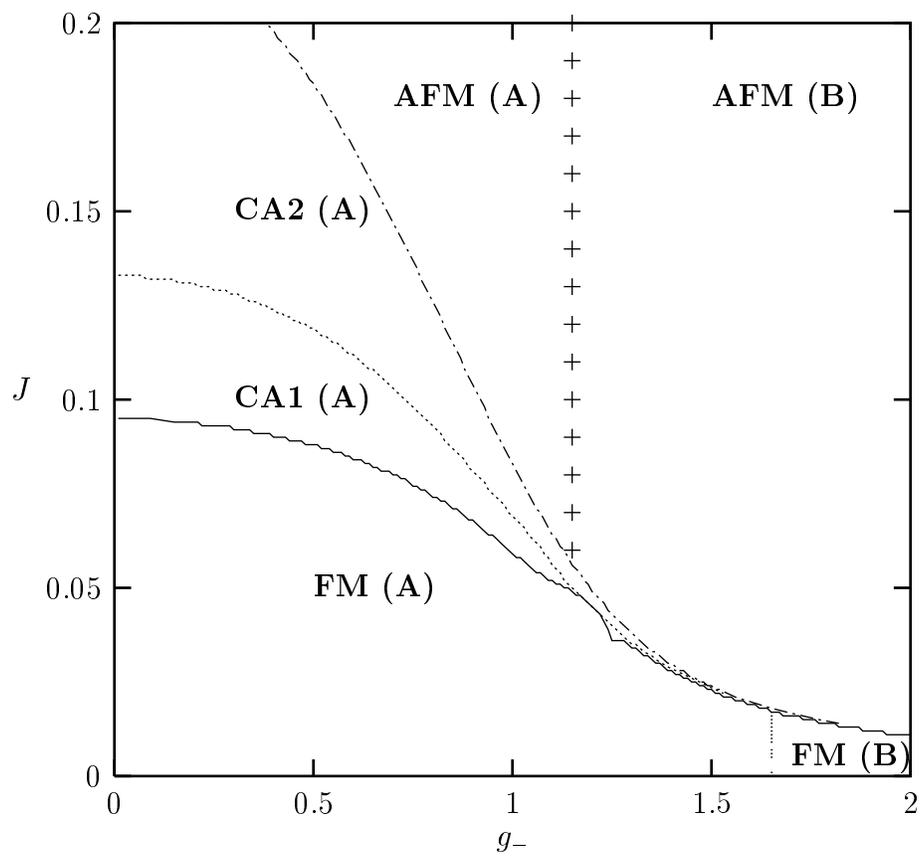
FIG. 1. The $g_{-}$ vs $J$ phase diagram ($\vec h = 0$) for 
$\mid \vec S_1 \mid =
\mid \vec S_2 \mid = \frac{3}{2}$ and $t=1$. ({\bf A}) and ({\bf B}) denote
large polaron and small polaron region respectively.

\vskip 0.5cm

\noindent 

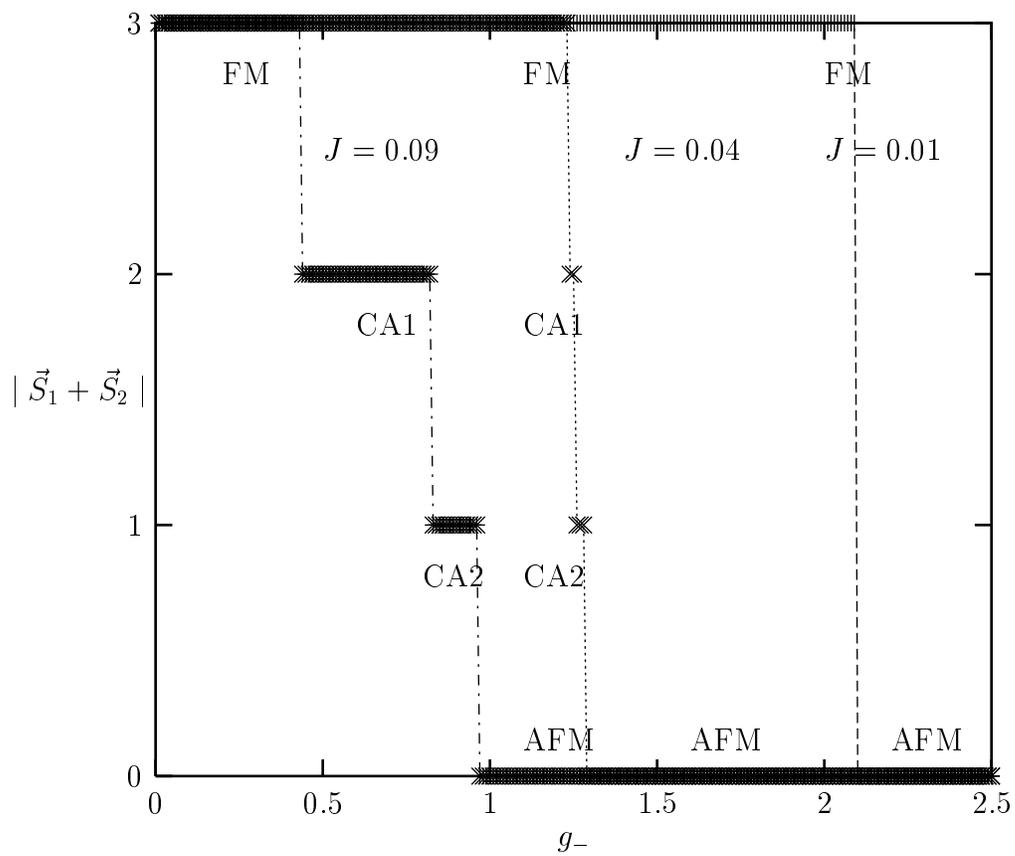
FIG. 2. Variations of ground state spin configuration 
$\mid \vec S_1 + \vec S_2 \mid$ with $g_{-}$ for $t=1$ and $J=$ 0.01, 0.04 and
0.09, $h=0$ (in units of $\omega_0 = 1$).

\vskip 0.5cm

\noindent

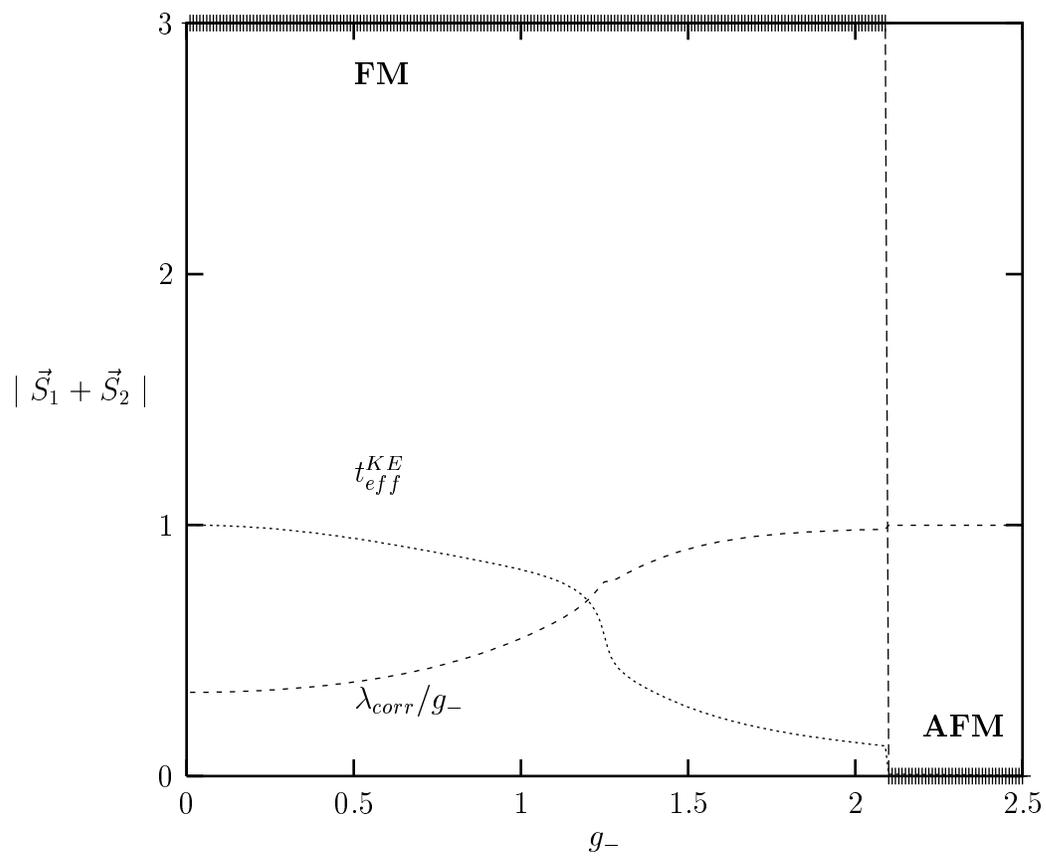
FIG. 3. Variations of $t_{eff}^{KE}$ and
$\lambda_{corr}/g_-$ with $g_-$ for $t=1.0$, $J=0.01$ and $h=0$.

\vskip 0.5cm

\noindent

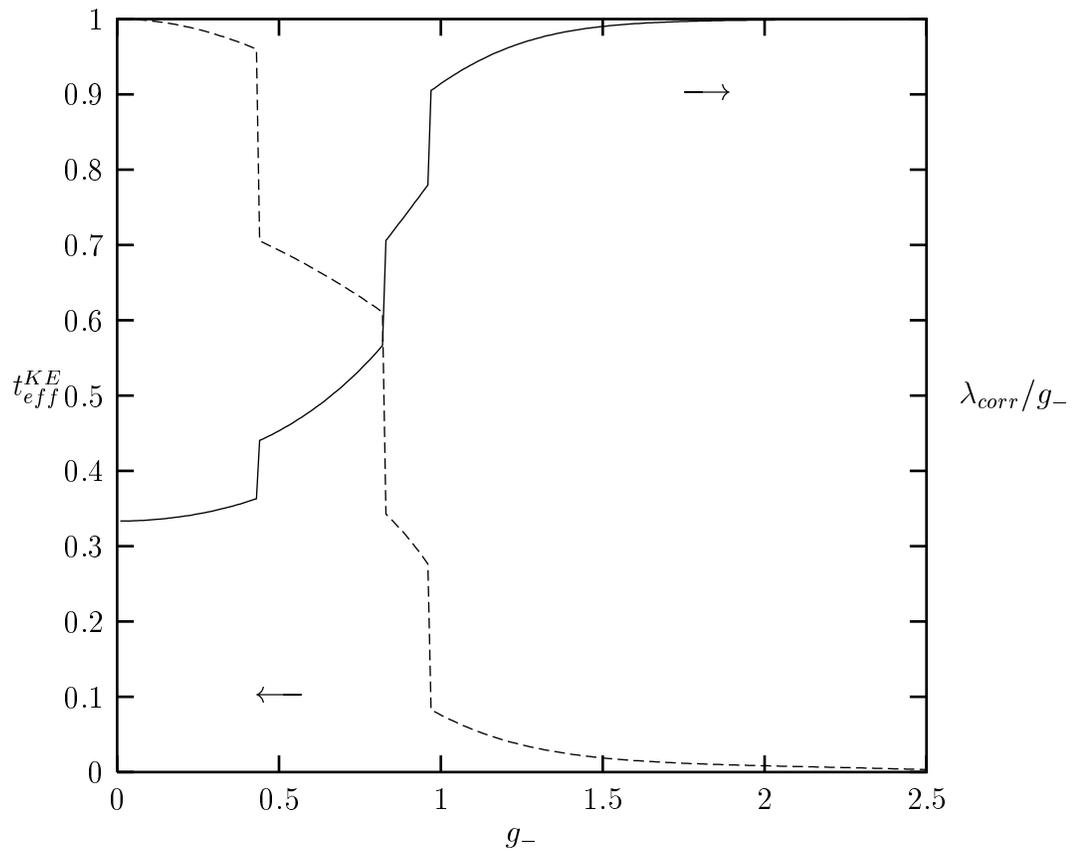
FIG. 4. Variations of effective kinetic energy $t_{eff}^{KE}$ (dashed line)
and polaron crossover $\lambda_{corr}/g_-$ (solid line) with $g_-$, 
for $t=1.0$, $J=0.09$ and $h=0$. 
The sharp jumps in $t_{eff}^{KE}$ and $\lambda_{corr}/g_-$ occur at values of 
$g_-$ where the magnetic transitions take place (see Fig. 2).

\vskip 0.5cm

\noindent

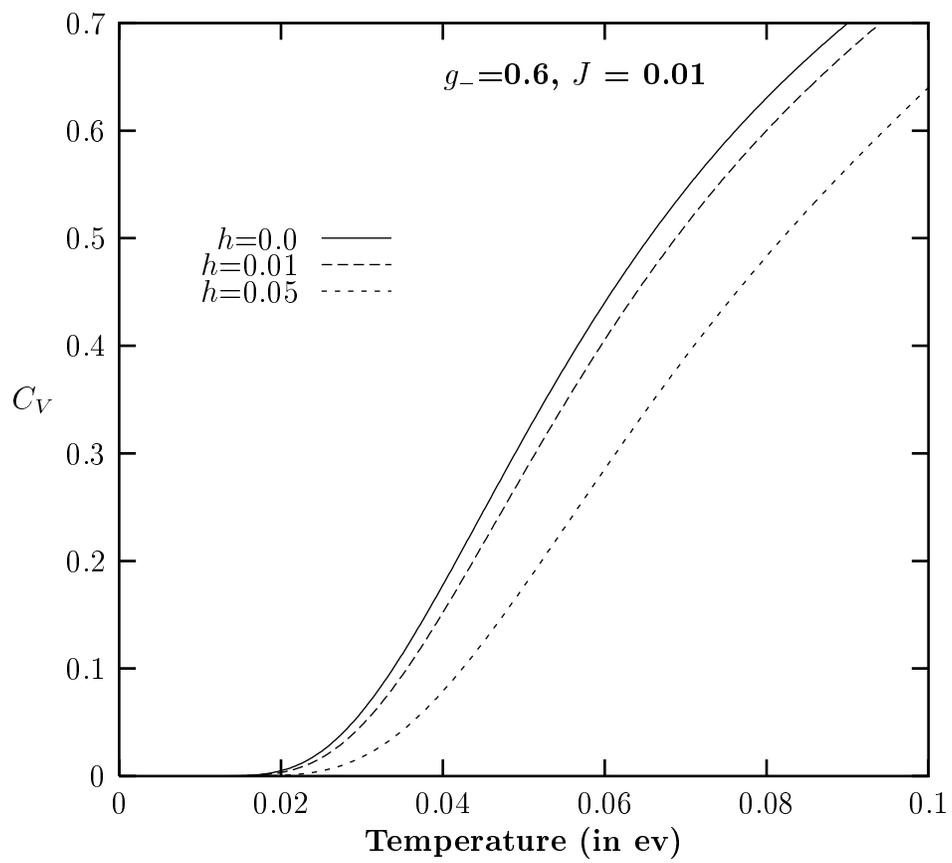
FIG. 5. Variations of $C_V$ (in arbitrary units) for $g_-=0.6$, $J=0.01$ and
$t=1$, for different values of the magnetic field $h=$ 0, 0.01, 0.05.

\noindent

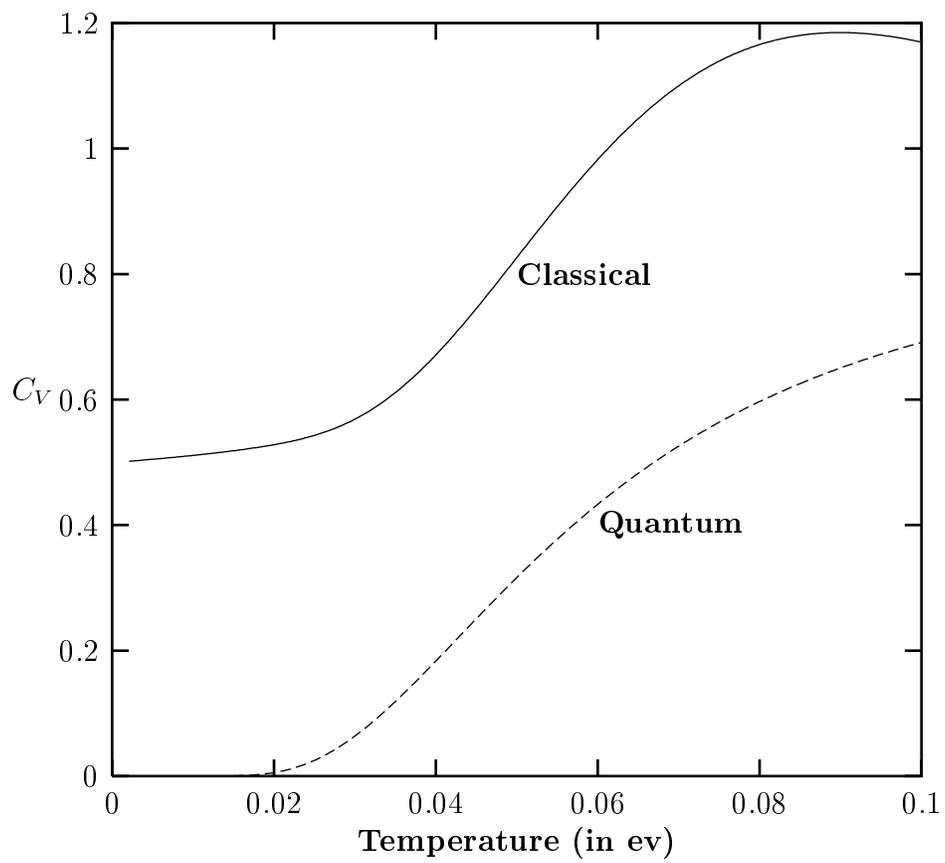
FIG. 6. 
Variations of $C_V$ (in arbitrary units) for $g_-=0.2$, $J=0.02$, $h=0$ and
$t=1$, in classical (solid line) and quantum (dashed line) formulation
of the core spins. The ground state is FM.
\noindent

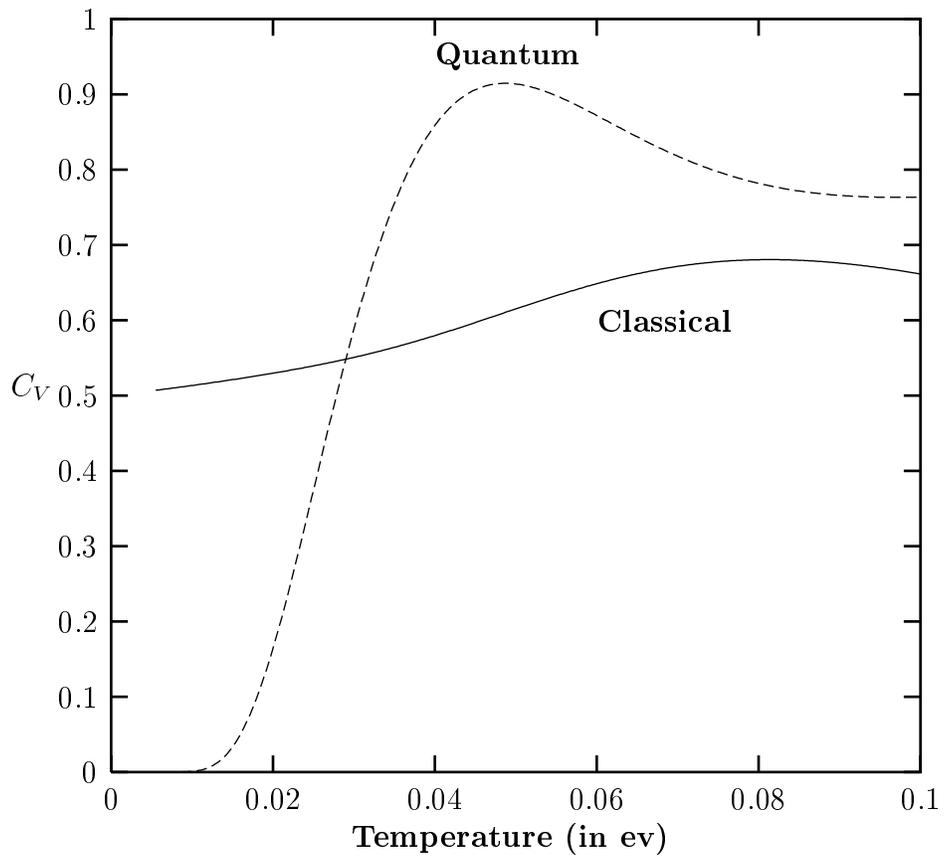
FIG. 7.
Variations of $C_V$ (in arbitrary units) for $g_-=0.9$, $J=0.2$, $h=0$ and
$t=1$, in classical (solid line) and quantum (dashed line) formulation
of the core spins. The ground state is AFM.


\begin{thebibliography}{999}

\bibitem{jin}S. Jin, T. Tiefel, M. McCormack, R. Fastnacht, R. Ramesh and
L. Chen, Science {\bf 264}, 413 (1994); S. Jin, M. McCormack, T. Tiefel and
R. Ramesh, Jour. Appl. Phys. {\bf 76}, 6929 (1994).

\bibitem{zen}C. Zener, Phys. Rev. {\bf 82}, 403 (1951)

\bibitem{and}P. W. Anderson and H. Hasegawa, Phys. Rev, {\bf 100}, 675 (1955). 

\bibitem{degen} P. G. de Gennes, Phys. Rev. {\bf 118}, 141 (1960).

\bibitem{kubo} K. Kubo and N. Ohata, Jour. Phys. Soc. Japan, {\bf 33}, 21 
(1972).

\bibitem{rev1} Myron B. Salamon and Marcelo Jaime, Rev. Mod. Phys. {\bf 73},
583 (2001).
\bibitem{rev2} J. M. D. Coey, M. Viret, S. von Molnar, Adv. in Phys. {\bf 48},
167(1999).

\bibitem{racah} L. C. Biedenharn, J. M. Blatt and M. E. Rose, Rev. Mod. Phys.
{\bf 24}, 249 (1952)

\bibitem{millis95} A. J. Millis, P. B. Littlewood and B. I. Shraiman,
Phys. Rev. Lett. {\bf 74}, 5144(1995). 

\bibitem{jayee}A. N. Das and Jayita Chatterjee, Int. Jour. Mod. Phys.,
{\bf 13}, 3903 (1999); Jayita Chaterjee and A. N. Das, Phys. Rev. B
{\bf 61}, 4592 (2000).

\bibitem{euro} J. Chatterjee, M. Mitra and A. N. Das, Euro. Phys. J. B 
{ \bf 18} , 573(2000). 

\bibitem{cv} T. Okuda. A. Asamitsu, Y. Tomioka, T. Kimura, 
Y. Taguchi and Y. Tokura, Phys. Rev. Lett. {\bf 81}, 3203(1998). 

\bibitem{cv1}T. Okuda, T. Kimura and Y. Tokura, Phys. Rev. B {\bf 60}, 
3370(1999).

\bibitem{cv2} M. Roy, J. F. Mitchell, A. P. Ramirez, P.
Schiffer, cond-mat/0101223.

\bibitem{sd} see, for example, L. P. Kadanoff and G. Baym,
\underline{Quantum Statistical Mechanics} (Benjamin, N. Y., 1962).


\end{thebibliography}
\end{document}